\documentclass[useAMS,usenatbib]{mn2e}

\usepackage{graphicx}

\title[Field T dwarfs in the UKIDSS GCS]{
Identifying nearby field T dwarfs in the UKIDSS Galactic Clusters Survey
\thanks{Based on observations made with the United Kingdom Infrared
Telescope, operated by the Joint Astronomy Centre on behalf of the
U.K. Science Technology and Facility Council.}}
\author[N. Lodieu et al.]{N. Lodieu$^{1}$\thanks{E-mail: nlodieu@iac.es},
B. Burningham$^{2}$, N. C. Hambly$^{3}$, D. J. Pinfield$^{2}$ \\
$^{1}$Instituto de Astrof\'isica de Canarias, V\'ia L\'actea s/n, 
E-38205 La Laguna, Tenerife, Spain \\
$^{2}$Centre for Astrophysics Research, Science and Technology Research 
Institute, University of Hertfordshire, Hatfield AL10 9AB \\
$^{3}$Scottish Universities' Physics Alliance (SUPA),
Institute for Astronomy, School of Physics and Astronomy, University of Edinburgh, \\
Royal Observatory, Blackford Hill, Edinburgh EH9 3HJ, U.K.  \\
}

\begin{document}

\date{Accepted \today. Received \today; in original form \today}

\pagerange{\pageref{firstpage}--\pageref{lastpage}} \pubyear{2005}

\maketitle

\label{firstpage}

%
%
\begin{abstract}
We present the discovery of two new late-T dwarfs identified in the 
UKIRT Infrared Deep Sky Survey (UKIDSS) Galactic Clusters Survey (GCS)
Data Release 2 (DR2). These T dwarfs are nearby old T dwarfs
along the line of sight to star--forming regions and open clusters
targeted by the UKIDSS GCS\@. They are found towards the $\alpha$\,Per
cluster and Orion complex,respectively, from a search in 54 square
degrees surveyed in five filters.
Photometric candidates were picked up in two--colour diagrams, 
in a very similar manner to candidates extracted from the UKIDSS 
Large Area Survey (LAS) but taking advantage of the $Z$ filter employed
by the GCS\@. Both candidates exhibit near--infrared $J$-band spectra with 
strong methane and water absorption bands characteristic of late--T 
dwarfs. We derive spectral types of T6.5$\pm$0.5 and T7$\pm$1 and
estimate photometric distances less than 50 pc for 
UGCS J030013.86$+$490142.5 and UGCS J053022.52$-$052447.4, respectively.
The space density of T dwarfs found in the GCS seems consistent with
discoveries in the larger areal coverage of the UKIDSS Large Area Survey,
indicating one T dwarf in 6--11 square degrees.
The final area surveyed by the GCS, 1000 square degrees in five
passbands, will allow expansion of the LAS search area  by 25\%,
increase the probability of finding ultracool brown dwarfs, and
provide optimal estimates of contamination by old field brown dwarfs
in deep surveys to identify such objects in open clusters and 
star--forming regions.
\end{abstract}


%
%
\section{Introduction}
\label{dT:intro}

The advent of large--scale sky surveys has revolutionised our knowledge
of ultracool dwarfs (refered to as dwarfs with spectral types of M7
and later). The first spectroscopic brown dwarfs were confirmed
in 1995: Gl229B, a T dwarf orbiting an M dwarf \citep{nakajima95}
and Teide~1 in the Pleiades open cluster \citep{rebolo95}.
More than twenty years later, about 530 L dwarfs with effective 
temperatures (T$_{\rm eff}$) between $\sim$2200 and $\sim$1400\,K 
\citep{basri00,leggett00} have now been identified (as of May 2008), 
along with around 140 T dwarfs with lower temperatures 
\citep[T$_{\rm eff}\simeq$ 1400--700 K;][]{golimowski04a,vrba04}.
The full catalogue of L and T dwarfs is available on the DwarfArchives.org
webpage\footnote{http://dwarfarchives.org,
a webpage dedicated to M, L, and T dwarfs maintained by C.\ Gelino, 
D.\ Kirkpatrick, and A.\ Burgasser.}.
There are currently 63 T5 or later dwarfs at the time of writing, refered to
here as late-T dwarfs. The spectral classification of T dwarfs follows the 
unified scheme by \citet{burgasser06a} and is based on the strength 
of methane and water absorption bands present in the near-infrared. 
This sample of ultracool dwarfs is now
large enough to characterise the binary properties of brown
dwarfs \citep{close02b,burgasser03a,bouy03,burgasser06d,liu06,burgasser07d} 
and investigate the influence of gravity and metallicity on their 
spectral energy distributions \citep{kirkpatrick05,jameson08b}.

The UKIRT Infrared Deep Sky Survey (UKIDSS) is defined in 
\citet{lawrence07}. UKIDSS uses the Wide Field Camera 
\citep[WFCAM;][]{casali07}
installed on the UK InfraRed Telescope (UKIRT) and a photometric system 
described in \citet{hewett06}. The pipeline processing is made at the
Cambridge Astronomical Unit (CASU; Irwin et al., in prep)\footnote{http://casu.ast.cam.ac.uk/surveys-projects/wfcam/technical} and the data
are available from the WFCAM Science Archive \citep{hambly08}.
The survey is now well underway with several ESO-wide releases: 
the Early Data Release (EDR) in February 2006 \citep{dye06} and the Data 
Release 1 \citep[DR1;][]{warren07a} in 2006, the Data Releases 2 \& 3
in March and December 2007 \citep{warren07b}\footnote{DR2 is now a public
worldwide release since September 2008}, and DR4 in July 2008,
the latest to date.

The Galactic Cluster Survey (GCS), one of the five components of UKIDSS, 
will cover 1000 deg$^2$ in $ZYJHK$ in ten clusters and star--forming 
regions down to a 5$\sigma$ sensitivity limit of $K\simeq$ 18.7 mag 
after two epochs at $K$. The main science goal of the GCS is to derive 
the very low--mass stellar/substellar mass function in
several open clusters and star--forming regions over large areas to
investigate the (possible) role of environment and time on the mass function.
We have already described the selection procedure of low--mass stars,
brown dwarfs, and planetary-mass members of the Upper Sco association
\citep{lodieu06,lodieu07a} and confirmed many photometric candidates as 
young L dwarf members \citep{lodieu08a}. 
Additionally, we have derived the mass function in the Pleiades down to 
30 Jupiter masses over 12 square degrees and estimated the photometric 
brown dwarf binary fraction \citep{lodieu07c}. 

Although the GCS is not sensitive to young T dwarfs at the distance of 
the targeted clusters, it is possible to look for nearer, old field T--type 
dwarfs in a similar manner to the LAS 
\citep{kendall07,warren07c,lodieu07b,pinfield08,burningham08}
which is beneficial both in the context of field and cluster brown
dwarf studies. 
It is a headline science goal of UKIDSS to find the 
so-called Y dwarfs: current searches have focused on the LAS but
ignored the GCS as a ground to uncover field T dwarfs. Ultimately,
the GCS will provide an additional 25\% coverage in five filters to
the 4000 deg${^2}$ planned by the LAS\@.

In this paper we describe the first search using the GCS DR2 in this 
context and report the discovery of a T6.5$\pm$0.5 and a T7$\pm$1
towards $\alpha$ Per and Orion, respectively. 
In Section \ref{dT_GCS:selection} we describe the photometric selection 
of late--T dwarfs from two--colour diagrams. 
In Section \ref{dT_GCS:properties} we present the near--infrared spectra 
for both targets, assign spectral types based on the unified classification 
scheme by \citet{burgasser06a}, and derive their main properties. 
In Section \ref{dT_GCS:contamination} we examine the issue of field
dwarf contamination 
towards young clusters and star--forming regions targeted by the GCS\@.
In Section \ref{dT_GCS:prospects} we discuss future prospects offered
by the GCS to extend this preliminary search for field T dwarfs and
further constrain levels of field dwarf contamination.

%
%
%
\begin{figure}
  \includegraphics[width=\linewidth, angle=0]{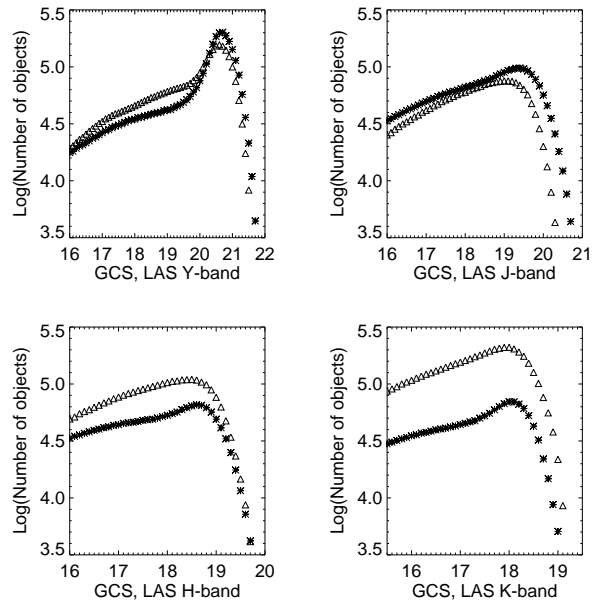}
   \caption{Histograms for the number of objects in the LAS (black)
and the GCS (red) per magnitude bin as a function of magnitudes for 
the $Y$ (top left), $J$ (top right), $H$ (bottom left), and $K$ 
(bottom right) filters. The 100\% completeness limits are quoted in
Table \ref{tab_dT_GCS:depth_compare}.
}
   \label{fig_dT:histograms}
\end{figure}
%

%
%
\section{Sample selection}
\label{dT_GCS:selection}
%


%
\subsection{The GCS coverage}
\label{dT_GCS:selection_coverage}

\begin{figure*}
\includegraphics[viewport = 100 0 443 790, width=0.45\linewidth, angle=270]{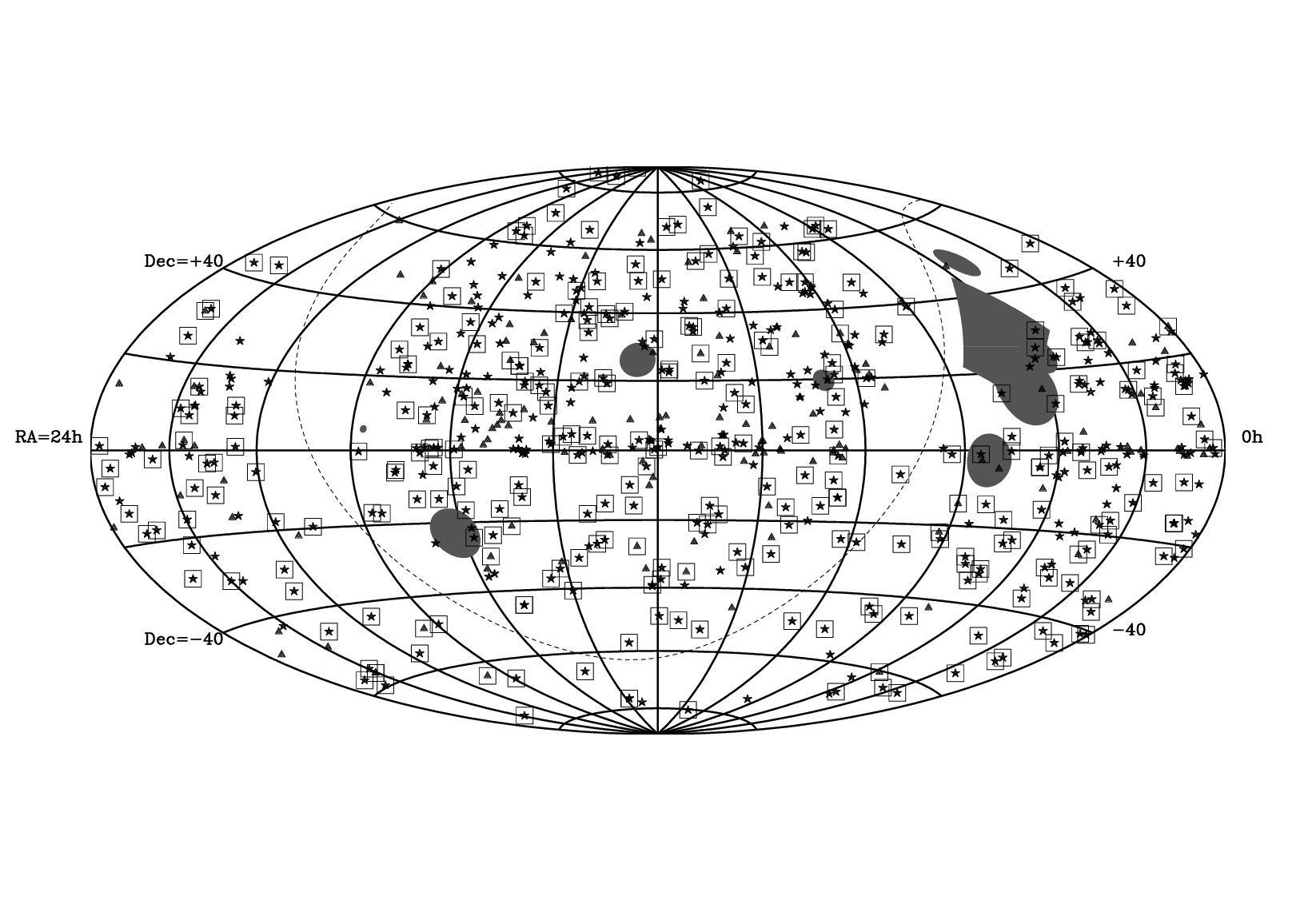}
\caption{Location of the known L (star symbols) and T (filled
triangles) dwarfs listed in the DwarfArchives.org webpage in the sky.
Open squares represent the brightest L and T dwarfs with $J \leq$ 15.5 mag.
These objects were essentially identified in 2MASS and SDSS\@.
We have excluded young L dwarfs members of the Pleiades \citep{martin99c},
$\sigma$ Orionis \citep{zapatero00}, and Upper Sco \citep{lodieu08a} that
were identified in deeper surveys.
Overplotted are the 10 open clusters and star--forming regions (grey areas)
that the GCS will fully cover in five filters after completion of the survey.
}
\label{fig_dT:GCScoverage}
\end{figure*}

The GCS is designed to cover fully 10 open clusters and star--forming 
regions. The Second Data Release provides coverage in many regions
at least in one filter, usually $K$. However,
only 54 square degrees have been surveyed in the Pleiades (8.8 deg$^{2}$), 
$\alpha$ Per (7.6 deg$^{2}$), Upper Sco (7.5 deg$^{2}$), 
IC\,4665 (4.05 deg$^{2}$), Taurus-Auriga (13.3 deg$^{2}$), and 
Orion (13 deg$^{2}$) and have been released 
in DR2\@. A cross--match between the L and T dwarfs listed in the 
dwarfarchives.org webpage$^{1}$ with the GCS DR2 returned only one source 
with a spectral type later than T5 (Fig.\ \ref{fig_dT:GCScoverage}):
2MASS J04070885$+$1514565 \citep[$J$ = 16.05 mag; T5;][]{burgasser04c}
towards the Hyades cluster. This object was only observed in the $K$-band
and thus was not recovered by our search.
The number density of T dwarfs (and more generally ultracool dwarfs)
varies across the sky (when measured in an aperture of a few degrees radius 
i.e.\ the size of a typical cluster). Hence, it is highly desirable to 
determine the number of such objects towards clusters if one is to 
accurately account for such contamination in cluster studies. 
The GCS offers a new opportunity to extend the search
for field brown dwarfs towards clusters and star--forming regions
to a significantly greater depth than 2MASS\@.

%
%
%
\subsection{LAS/GCS comparison}
\label{dT_GCS:selection_comparison}

In addition to the traditional $JHK$ filters and the $Y$ filter centered 
at 1.03 (0.98--1.08) microns used by the LAS, the GCS also employs
a $Z$ filter centered at 0.88 (0.83--0.925) microns. We will take advantage 
of this filter to select old T dwarfs and effectively replace the $z$-band 
filter from the Sloan Digital Sky Survey (SDSS) that is used to search 
for late--T dwarfs in the LAS \citep[e.g.][]{lodieu07b}. Note that the GCS 
exposure times and quality control procedures are the same as those for the 
LAS for the filters in common. This has been done to enable homogeneous 
sample selections across both surveys.

We have compared the depth of the Second Data Release of the LAS and
GCS \citep{warren07b} for the four common filters ($YJHK$).
Figure \ref{fig_dT:histograms} displays the histograms of the number
of objects as a function of magnitude for the LAS and GCS surveys.
Completeness limits for the four filters are very similar for both
surveys as expected from the similar exposure times although the
GCS tends to be slightly shallower (Table \ref{tab_dT_GCS:depth_compare}).

We have also compared the effects of optical-to-infrared colour selection:
we used the ($Z-J$)$_{\rm UKIDSS}$ for the GCS as opposed to the
$z_{\rm SDSS}-J_{\rm GCS}$ for the LAS\@. T dwarfs have typically
$z_{\rm SDSS}-J_{\rm GCS}$ redder than 2.6 mag \citep{hawley02}
criterion used for searches in the LAS\@. \citet{hewett06} report
synthetic colours for T dwarfs of ($Z-J$)$_{\rm GCS}$ redder than
2.8 mag. Therefore, our colour criteria defined in 
Sect.\ \ref{dT_GCS:selection_procedure} should pick up all T dwarfs 
that the LAS criteria would identify. However, the rejection rate
of contaminating sources might be lower at the faint end because
the $Z_{\rm GCS}$ detection limit is shallower than the $z_{\rm SDSS}$ 
limit (20.4 mag vs 20.8 mag).
sources.

%
%
%
\begin{table}
 \centering
 \caption[]{Comparison of the depth between the LAS and GCS: the
numbers quoted correspond to the point where the number of objects
as a function of magnitude deviates from a straight line.
Values are given for the 2$^{\rm nd}$ Data Release of the LAS and GCS\@.
}
 \begin{tabular}{l c c c c}
 \hline
Survey &  $Y$  &  $J$  &  $H$  & $K$  \cr
\hline
LAS    &  20.7 &  19.4 & 18.6  & 18.0 \cr
GCS    &  20.6 &  18.9 & 18.3  & 17.9 \cr
\hline
 \label{tab_dT_GCS:depth_compare}
 \end{tabular}
\end{table}
%

%
%
%
\begin{table*}
 \centering
 \caption[]{Coordinates (J2000), infrared magnitudes
            ($YJHK$) on the WFCAM MKO system with their associated
            errors taken from the Second Data Release of the GCS
            \citep{warren07b}, spectral types with uncertainties,
            and estimated photometric distances for both T dwarfs
            identified in the UKIDSS GCS and confirmed spectroscopically
            with Gemini/NIRI\@.}
 \begin{tabular}{c c c c c c c c c}
 \hline
Name  &  RA   & dec   &   $Y$  &  $J$  &  $H$  & $K$  & SpType & dist \cr
\hline
UGCS J030013.86$+$490142.5  & 03 00 13.86 & $+$49 01 42.5 & 18.99$\pm$0.07 & 17.99$\pm$0.04 & 18.37$\pm$0.16 & $>$18.2 & T6.5$\pm$0.5 & 20--46 pc \cr
UGCS J053022.52$-$052447.4  & 05 30 22.52 & $-$05 24 47.4 & 19.45$\pm$0.12 & 18.35$\pm$0.07 & 19.13$\pm$0.29 & $>$18.2 & T7.0$\pm$1.0 & 19--42 pc \cr
\hline
 \label{tab_dT_GCS:objects}
 \end{tabular}
\end{table*}
\subsection{Selection procedure}
\label{dT_GCS:selection_procedure}

Figure 1 of \citet{lodieu07b} displays a ($Y-J$,$J-H$) two-colour diagram
showing the location of spectroscopically confirmed field L and T dwarfs 
from UKIDSS DR1 \citep{warren07a}, along with model predictions
\citep{allard01, marley02,baraffe03,tsuji04}.
T dwarfs exhibit neutral to blue near-infrared
colours ($J-H$ or $J-K$) with decreasing effective temperature 
\citep{burgasser02} and a $Y-J$ colour between 0.7 and 1.3 mag with
a possible trend towards a bluer $Y-J$ colour with cooler temperatures
\citep{pinfield08}. On the other hand, SDSS discoveries 
\citep{geballe02,hawley02} show that T dwarfs have red optical-to-infrared 
colours (typically $z-J \geq 2.5$). Furthermore, atmosphere models predict 
blue near-infrared colours ($J-H < 0.0$) for late-T dwarfs and cooler objects
\citep{kirkpatrick99}. However, their predictions differ on the expected 
$Y-J$ as a function of temperature with a colour range between 0.5 and 
1.5 mag.

\begin{figure}
  \includegraphics[width=\linewidth, angle=0]{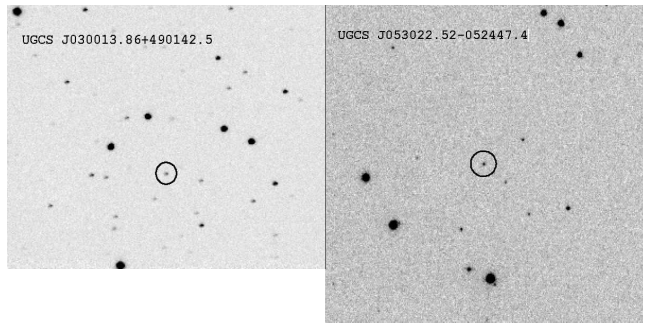}
   \caption{Finding charts for the two new nearby old T dwarfs
extracted from the UKIDSS GCS DR2: UGCS0300 (left) and UGCS0530 (right).
Finding charts are $J$-band images of 2 arcmin on a side with North up
and East left.
}
   \label{fig_dT:fcTs}
\end{figure}

We have applied the following selection constraints to look for old field 
late-T dwarfs in the 54 square degrees surveyed in $ZYJHK$ by the GCS DR2 
\citep{warren07b}:
\begin{itemize}
\item $Y-J \geq$ 0.7 mag
\item $J-H \leq-$0.2 mag
\item Z non detection or $Z-Y \geq$1.7 mag 
\item $J$ = 15.0--19.6 mag and $H \leq$ 19.2 mag
\item Point sources i.e.\ {\tt{mergedClass}}$^2$ = $-$1
\item Good quality detections i.e.\ {\tt{ppErrBits}}\footnote{Definitions
of these parameters can be found in the web pages of the WFCAM Science 
Archive}$\leq$ 256
\end{itemize}
The query returned three potential late--T dwarfs and we present here 
$J$--band spectroscopy for two of them (Table \ref{tab_dT_GCS:objects};
Figs.\ \ref{fig_dT:fcTs} \& \ref{fig_dT_GCS:Tdwarfs}), 
UGCS J030013.86$+$490142.5 (hereafter UGCS0300) 
and UGCS J053022.52$-$052447.4 (UGCS0530). 
We note that no bright sources detected in $Z$ and satisfying 
$Z-Y \geq$1.7 mag were identified.

%
%
%
\section{Properties of the new T dwarfs}
\label{dT_GCS:properties}
\subsection{Gemini/NIRI spectroscopy}

Spectroscopy in the $J$-band was obtained using the Near InfraRed
Imager and Spectrometer \citep[NIRI;][]{hodapp03} on the Gemini North
Telescope on Mauna Kea under program GN-2007B-Q-26\@.
UGCS0300 was observed on the night of 2007 December 23, and UGCS0530
was observed on the night of 2008 January 04\@.
All observations were made up of a set of sub-exposures in an ABBA
jitter pattern to facilitate effective background subtraction, with a
slit width of 1 arcsec. The length of the A-B jitter was 10 arcsecs.
The total integration time for UGCS0300 was 16 minutes, whilst for
UGCS0530 it was 32 minutes.
The observations were reduced using standard IRAF Gemini packages
\citep{lodieu07b,pinfield08}.

A comparison argon arc frame was
used to obtain a dispersion solution, which was then applied to the
pixel coordinates in the dispersion direction on the images.
The resulting wavelength-calibrated subtracted pairs had a low-level
of residual sky emission removed by fitting and subtracting this
emission with a set of polynomial functions fit to each pixel row
perpendicular to the dispersion direction, and considering pixel data
on either side of the target spectrum only.
The spectra were then extracted using a linear aperture, and cosmic
rays and bad pixels removed using a sigma--clipping algorithm.

Telluric correction was achieved by dividing each extracted target
spectrum by that of an F type star observed just before or after the 
target and at a similar airmass.
Prior to division, hydrogen lines were removed from the standard star
spectrum by interpolating the stellar
continuum.
Relative flux calibration was then achieved by multiplying through by a
blackbody spectrum of the appropriate T$_{\rm eff}$.
The spectra were then rebinned by a factor of three to increase the
signal-to-noise, whilst avoiding under-sampling for the spectral resolution.
final spectra are shown in Fig.\ \ref{fig_dT_GCS:Tdwarfs}.

%
%
%
\subsection{Spectral classification}
\label{dT_GCS:classification}

\begin{figure}
  \includegraphics[width=\linewidth]{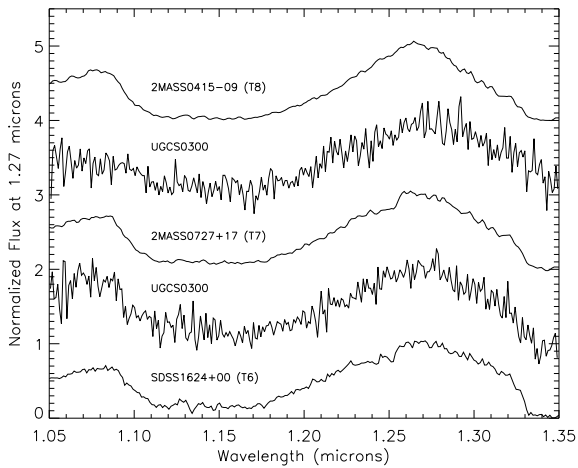}
   \caption{Gemini/NIRI $J$-band (1.05--1.35 microns) spectra
(binned by a factor three) of two new T dwarfs identified in
the UKIDSS GCS (UGCS0300 and
UGS0530). Overplotted are spectral templates for direct comparison,
including SDSSp J162414.37$+$002915.6 \citep[T6;][]{strauss99},
2MASSI J0727182$+$171001 (T7), and
2MASSI J0415195$-$093506 \citep[T8;][]{burgasser02,burgasser06a}.
}
   \label{fig_dT_GCS:Tdwarfs}
\end{figure}

Figure \ref{fig_dT_GCS:Tdwarfs} compares the spectra for UGCS0300 and 
UGCS0530 with three spectral standards:
SDSSp J162414.37$+$002915.6 \citep[T6;][]{strauss99}, 
2MASSI J0727182$+$171001 (T7) and 
2MASSI J0415195$-$093506 \citep[T8;][]{burgasser02,burgasser06a}.
Although the $Y$-band peak appears to be enhanced in UGCS0300 with
respect to the spectral type standards, this comparison suggests a
spectral type of T6.5 (Fig.\ \ref{fig_dT_GCS:Tdwarfs}).
The calculated spectral indices on the system of \citet{burgasser06a}
and \citet{burningham08} are given in Table \ref{tab_dT_GCS:SpTypes}.
These tend to support the spectral type implied by template
comparison, and we adopt a spectral type of T6.5$\pm 0.5$ for
UGCS0300\@.

The template comparison across most of the spectral coverage suggests
a T7 classification for UGCS0530\@. The red side of the J-band peak, 
however, is at odds with this, appearing brighter and reminiscient of 
an earlier type. The calculated spectral indices in 
Table \ref{tab_dT_GCS:SpTypes} (CH$_4$--J index) reflect this morphology.
Considering the low signal of the spectrum in this region, we choose
to disregard it in this classification, pending higher signal-to-noise
observations with broader coverage at a later date. We adopt a type of
T7$\pm$1 for UGCS0530\@.

%
%
%
\begin{table}
\caption{Spectral types for UGCS0300 and UGCS0530 derived from
direct comparison with template T dwarfs, and several
spectral indices defined in the literature.}
 \begin{tabular}{c c c}
  \hline
Indicator  &  UGCS0300 & UGCS0530 \\
  \hline
Template & T6.5$\pm$0.5 & T7$\pm$1 \\
  \hline
H$_2$O--J & 0.145$\pm$0.015 (T6)   & 0.108$\pm$0.018 (T7) \\
CH$_4$--J & 0.212$\pm$0.014 (T7)   & 0.537$\pm$0.017 (T3) \\
$W_J$     & 0.39$\pm$0.01 (T7/T6)  & 0.40$\pm$0.01 (T6/T7)\\
  \hline
Adopted type & T6.5$\pm 0.5$ & T7$\pm 1$\\
\hline
  \label{tab_dT_GCS:SpTypes}
 \end{tabular}
\end{table}

%
%
%
\subsection{Distances of the new T dwarfs}
\label{dT_GCS:distances}

\begin{figure}
  \includegraphics[width=\linewidth]{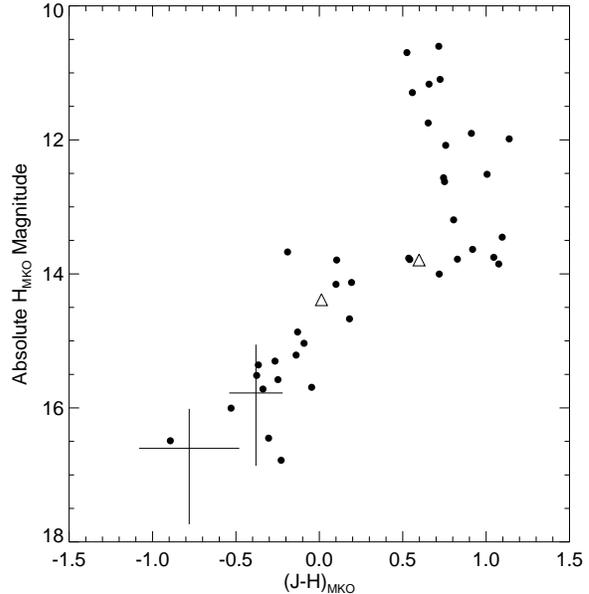}
   \caption{$J-H$ colour as a function of absolute $H$ magnitude
in the MKO system. The two new T dwarfs identified in the GCS
are marked with crosses representing the errors in the colours
and distance estimates. For comparison, we have also plotted
two Pleiades T dwarfs candidates (not confirmed spectroscopically
yet) as open triangles \citep{casewell07}.
}
   \label{fig_dT_GCS:colour_compare}
\end{figure}

We have identified two T dwarfs with spectral types later than T5 in
54 square degrees surveyed in all five filters by the GCS DR2\@.
Those numbers are consistent with the eight late-T dwarfs found
in the LAS DR2 over 282 square degrees surveyed in $YJHK$ and
satisfying the criteria described in Sect.\ \ref{dT_GCS:selection}.
Since these are field T dwarfs and not cluster T dwarfs, we can
simply estimate their photometric distances by comparison with known
T dwarfs with measured parallaxes
\citep[Fig.\ \ref{fig_dT_GCS:colour_compare};][]{vrba04,tinney03}.

There are three known T6.5 dwarfs (infrared spectral types based on the
classification of \citet{burgasser06a}) listed in the brown dwarf archive
with measured parallaxes and not part of a binary system:
2MASSI J104753.8$+$212423 ($J$ = 15.819 mag and d = 10.6 pc),
2MASSI J123739.2$+$652614 \citep[$J$ = 16.053 mag and d = 10.4 pc;][]{burgasser99,vrba04},
and SDSSp J134646.4$-$003150 \citep[$J$ = 16.0 mag and d = 14.6 pc;][]{tsvetanov00,tinney03}.
Assuming a spectral type of T6.5 for UGCS0300 and assuming that the
object is single, we derive an average distance of 30 pc with a 
possible range of 25--37 pc (1$\sigma$ limits). We estimate an
error on the distance of $\sim$10 pc assuming a dispersion of $\pm$0.5 mag
on the spectral type-absolute magnitude relation, yielding an expected
distance interval of 20 to 46 pc.

Similarly, only one T dwarf with measured parallax has a spectral type
of T7 \citep{burgasser06a}, 2MASSI J072718.2$+$171001
\citep[$J$ = 15.6 mag and d = 9.08 pc;][]{burgasser02,vrba04}.
Assuming a spectral type of T7 for UGCS0530 and assuming that the
object is single, we derive a distance of 32 pc. We estimate an
error on the distance of $\sim$13 pc, mainly due to the uncertainty
on the spectral type ($\pm$1 subclass).

%
%
%
%

%
%
%
\begin{table*}
 \centering
 \caption[]{Expected numbers of T dwarf (T0--T8.5) contaminants 
(second lines) as a function of areal coverage in square degrees 
(first lines) in each individual young cluster 
and star--forming region targeted by the UKIDSS Galactic Cluster 
Survey. We have also included the expected numbers of late-T dwarfs
in 2MASS (2M), assuming a search complete to $J$ = 15.65 mag.
}
 \begin{tabular}{@{\hspace{0mm}}l c c c c c c c c c c c@{\hspace{0mm}}}
 \hline
      &  Total    & Pleiades   & $\alpha$\,Per & IC\,4665 & USco   & Coma Ber   & Praesepe   & Taurus     & Hyades     & Orion      & Per OB     \cr
\hline
2M    & 1000     & 79         & 50         & 4        & 154        &  79        & 28         & 218        &  291       & 154        & 13         \cr
      & 0.9--1.5 & 0.07--0.12 & 0.04--0.08 & $<$0.01  & 0.14--0.24 & 0.07--0.12 & 0.02--0.04 & 0.20--0.33 & 0.26--0.45 & 0.14--0.24 & 0.01--0.02 \cr
DR2   & 54       & 8.81       & 7.60       & 4.05     & 7.46       &  0         & 0          & 13.3       &  0         & 13.0       & 0          \cr
      & 5--9     & 0.8--1.4   & 0.7--1.2   & 0.4--0.  & 0.7--1.2   &  0         & 0          & 1.2--2.1   &  0         & 1.2--2.0   & 0          \cr
Full  & 1000     & 79         & 50         & 4        & 154        &  79        & 28         & 218        &  291       & 154        & 13         \cr
      & 93--157  & 7.3--12.4  & 4.6--7.9   & 0.4--0.6 & 14.3--24.2 & 7.3--12.4  & 2.6--4.4   & 20.2--34.2 & 27.0--45.7 & 14.3--24.2 & 1.2--2.1   \cr
\hline
 \label{tab_dT_GCS:nb_expected}
 \end{tabular}
\end{table*}
\subsection{Cluster membership}
\label{dT:cluster_memb}

Since the primary aim of the GCS is to test the universality of the 
substellar Initial Mass Function \citep[IMF;][]{salpeter55}, its emphasis 
is on achieving full spatial coverage of several young clusters and 
star--forming regions, rather than achieving the required depth to identify 
very faint T dwarfs within them. We have recently demonstrated that the
coolest brown dwarf candidates identified in Upper Sco are
early-L dwarfs \citep{lodieu08a}. Much deeper, but smaller area, surveys
of the Pleiades \citep{casewell07} and $\sigma$Ori
\citep{zapatero00,zapatero02b,martin03b,burgasser04b,zapatero08a}
have been able to extract candidate T dwarf members.


Figure \ref{fig_dT_GCS:colour_compare} compares the colours of UGCS0300 
and UGCS0530 with those of known single T dwarfs extracted from the 
L and T dwarfs archive$^{1}$ in the Mauna Kea Observatory system
\citep[MKO;][]{tokunaga02}. Both objects lie in the sequence of
late-T dwarfs and do not show obvious sign of peculiar colours that
could indicate a young age \citep{casewell07,zapatero08a}. 
The errors, represented by large crosses in 
Fig.\ \ref{fig_dT_GCS:colour_compare} come from the uncertainty on
the colour (GCS photomery) and from the uncertainty on the 
distance estimates (see Section \ref{dT_GCS:distances}).

Additionally, we have looked into the expected number of late-T 
dwarfs from the simulations published by \citet{caballero08e}.
These authors used a ``standard'' model for the Galaxy and compiled
available data on late-type dwarfs to estimate their absolute
magnitudes, colours, space densities, and scale heights.
For late-T dwarfs, they used a space density of 
1--5$\times$10$^{-3}$ pc$^{-3}$ based on a mass function defined
by dN/dM$\propto$M$^{-0.5}$ and a normalisation factor of 0.0037 pc$^{-3}$ 
over the 0.09--0.10 M$_{\odot}$ mass range \citep{reid97c,deacon08a}.
We have assumed that late-T dwarfs have typical $I-J$ colours
around 5 mag and that our search is complete down to $J$ = 19 mag
(Sect.\ \ref{dT_GCS:selection_procedure}).
Following Table 4 of \citet{caballero08e}, we have found that the
number of T5--T8 dwarfs in 54 square degrees range from 4.6 to 23
with a most probable value of nine. Our GCS search contains a
few additional potential T dwarfs not yet confirmed spectroscopically, 
suggesting that our results are most likely consistent with the 
lowest numbers predicted by \citet{caballero08e}. 
The difference between our numbers and the ones in \citet{caballero08e}
might originate from one of the assumptions on the properties (binary
fraction, mass ratio distribution, number density) of
late-T dwarfs made by \citet{caballero08e} or on the selected mass 
function. However, one would need to test the level of influence that
each (assumed) parameters has on the final numbers quoted by
\citet{caballero08e} to provide a statistical interpretation on
the observed difference. Note that testing the birth rate would require
a significantly larger number of T dwarfs, not yet available from
UKIDSS \citep{deacon08a}.

Larger areas will become available in upcoming GCS releases and should 
allow us to constrain further the level of contamination by late-T dwarfs 
towards clusters.

%
%
\section{Contamination towards clusters}
\label{dT_GCS:contamination}

We have confirmed spectroscopically two late-T dwarfs with $Y-J \geq$ 0.7 
mag, $J-H \leq -$0.2 mag, and $H \leq$ 19.2 mag 
(Sect.\ \ref{dT_GCS:selection_procedure}).  
These numbers are consistent with the number of late-T dwarfs identified 
in the LAS \citep{pinfield08} based on a larger sample and greater areal 
coverage. In total, \citet{pinfield08} estimated to 26--44 T0--T8.5 dwarfs 
in 280 square degrees taking  into account biases due to incompleteness 
and binarity, implying one T0--T8 dwarf per 6--11 deg$^{2}$ or 1.5 late-T 
dwarf ($\geq$ T4) in 54 deg$^{2}$. We can apply these values to the GCS 
because the LAS and the GCS share the same design \citep{lawrence07} and 
have comparable completeness limits \citep[Sect.\ \ref{dT_GCS:selection_comparison}; Table \ref{tab_dT_GCS:depth_compare};][]{warren07b}.

These new T dwarfs lie towards $\alpha$ Per, located at a distance of 
180 pc \citep{pinsonneault98,robichon99} and along the line of sight of 
Orion \citep[$\sim$480 pc;][]{genzel81}. They were identified in 7.6 
and 13 deg$^{2}$ coverage in $\alpha$ Per and Orion, respectively.
Those numbers match well the one T dwarf expected in 6--11 deg$^{2}$
of the LAS\@. Considering the area covered in Taurus, the Pleiades,
and Upper Sco, which are on the same order as the ones surveyed in 
$\alpha$ Per and Orion, we should have found one T dwarf along their
line of sight. However, this lack of detection likely originates from
our stringent constraints (Sect.\ \ref{dT_GCS:selection_procedure})
because neither late-T dwarfs with redder $J-H$ colours nor early-T dwarfs
are considered by our search. 

Table \ref{tab_dT_GCS:nb_expected} shows 
the expected numbers of T dwarfs (T0--T8.5) in the area surveyed in each
regions and released in the GCS DR2 as well as those expected after 
completion of the GCS\@. The latter values are approximate due to 
unpredictable factors that could affect the survey (e.g., quality control).
Those numbers are also compared to the expected numbers of T dwarfs
in each regions, assuming a coverage equivalent to the one after
completion of the GCS\@. Additionally, we have assumed a 5$\sigma$
completeness limit of $J$ = 16.55 and $H$ = 15.85 for 2MASS
\citep{cutri03}, corresponding to $J$ 15.65 mag to match our
search criteria (Sect.\ \ref{dT_GCS:selection_procedure}).
Repeating this exercise for L dwarfs based on known absolute
magnitudes and densities \citep{vrba04,caballero08e}, we should
expect around 380 L0--L8 dwarfs in the full 2MASS database.
The GCS should provide 100 times more L and T dwarfs, hence 
decreasing the uncertainty on estimates of the level of contamination.

More specifically, the Pleiades survey described in \cite{lodieu07c},
covering 12 square degrees would contain one T dwarf contaminant down
to its completeness limit of $J$ = 18.9 mag. Similarly, \citet{casewell07}
reported four T dwarf candidates in a 2.5 deg$^{2}$ area down to a
completeness limit of $J$ = 19.7 mag. Scaling the numbers derived
above, we argue that $\sim$1 of their candidates could be a field
T dwarf rather than a cluster T dwarf of 10 Jupiter masses.

Finally, we can compare our results to the simulations presented by
\citet{caballero08e}. Their work is based on a galactic model and
uses the most recent data available for field L and T dwarfs.
Assuming a mean $I-J$ colour of 5 for T0--T8 dwarfs (their Table 3)
and comparing the $\Delta$I magnitude range given in their Table 4,
the GCs is able to find T dwarfs down to $I$ = 24 mag. Adding up
all the numbers, \citet{caballero08e} calculated 0.218--0.823
T0--T8 dwarfs down to $I$ = 24 mag (or $J \sim$ 19 mag) in one
square degree, indicating a total of 1--8 T0--T8 dwarfs in 
6--11 deg$^{2}$. Their lower limit is consistent with our observational
results whereas their upper limits seem to overestimate significantly
the expected number of T dwarf contaminants.

%
%
%
\section{Future prospects}
\label{dT_GCS:prospects}

Our results, combined with our previous published studies in the LAS, show 
that the GCS is not only powerful in identifying young low--mass stars and 
brown dwarfs in clusters but also in finding nearby old late--T dwarfs in 
the field. Also, thanks to its large coverage, the GCS constitutes a tool
to constrain observationally the level of contamination at the faint end
of the cluster mass functions. Besides the natural extension of this work 
to upcoming UKIDSS releases and ultimately the full GCS, this work 
opens several advantages:

\begin{itemize}
\item Extend the search area of the LAS by 25\%, expand the sample of late--T 
dwarfs, and increase the chance to find Y dwarfs, one of the UKIDSS main 
science objectives. The LAS will cover 4000 square degrees in $YJHK$ while 
the GCS will survey 1000 square degrees in the same set of filters with 
an additional $Z$ passband acting as a proxy for the SDSS $z$ filter. 
Moreover, the contamination 
by young T dwarfs members of open clusters should be negligible since the GCS 
is not sensitive to detect such cool, dim brown dwarfs at the distance of
the various clusters/star--formation regions targeted by the GCS
(Sect.\ \ref{dT:cluster_memb}).
\item Pin down the level of contamination towards open clusters and 
star--forming regions where L and T dwarfs are now found with a wide 
range of ages: Upper Sco \citep{lodieu08a}, $\sigma$ Orionis 
\citep{zapatero00,caballero07d}, Alpha Per \citep{jameson08b}, the Pleiades 
\citep{bihain06,lodieu07c,casewell07}, Ursa Major and Hyades 
\citep{jameson08a}. Those numbers will likely grow quickly in the 
coming years with the advent of deeper and larger-scale sky 
surveys, including VISTA (Visible and Infrared Survey Telescope for 
Astronomy), PanStarrs, and the LSST (Large Synoptic Survey Telescope).
\item Expand the current search to field L dwarfs to provide an estimate 
of the level of contamination in deep surveys targeting open clusters and 
star--forming regions (references therein). This search will be affected 
by the presence of young L dwarfs belonging to the clusters targeted by 
the GCS because the $Y-J$ and $J-K$ colours get redder with younger ages 
\citep{jameson08b}.
\end{itemize}
%

%
%
\section{Conclusions}
\label{dT:conclusions}

We have presented the spectroscopic confirmation of two new nearby
old late--T dwarfs identified in 54 square degrees surveyed as part
of the Second Data Release of the UKIDSS Galactic Cluster Survey. 
Those sources lie in front the Alpha Per cluster and Orion complex 
at distances less than 50 pc. These sources have been identified
in regions previously ignored by earlier searches for cool brown dwarfs.

This work opens new prospects to (i) extend the search area of the
UKIDSS LAS and increase the number of late--T dwarfs by 25\%, and
(ii) estimate the contamination in deep optical and near--infrared
surveys of young open clusters and star--forming regions by nearby
old L and T dwarfs and thereof correct the faint end of the photometric
mass functions.

%
%
\section*{Acknowledgments}

This research has made use of the Simbad database of NASA's
Astrophysics Data System Bibliographic Services (ADS), and
has benefitted from the M, L, and T dwarf compendium housed
at DwarfArchives.org and maintained by Chris Gelino, Davy Kirkpatrick,
and Adam Burgasser.

The United Kingdom Infrared Telescope is operated by the Joint 
Astronomy Centre on behalf of the U.K.\ Science and Technology Facilities 
Council. 

Based on observations obtained at the Gemini 
Observatory (program GN--2007B--Q--26), which is operated by the 
Association of Universities for Research in Astronomy, 
Inc., under a cooperative agreement with the NSF on behalf of the 
Gemini partnership: the National Science Foundation (United States), 
the Science Technology and Facility Council (United Kingdom), 
the National Research Council (Canada), CONICYT (Chile), the Australian 
Research Council (Australia), CNPq (Brazil) and SECYT (Argentina).

%
%
\bibliographystyle{mn2e}
\bibliography{../../AA/mnemonic,../../AA/biblio_old}

\label{lastpage}

\end{document}